\documentclass[traditabstract,letter,longauth]{aa}  
\usepackage[authoryear]{natbib}
\usepackage{graphicx}

\def\gtrsim{~\rlap{$>$}{\lower 1.0ex\hbox{$\sim$}}}

\begin{document}

\title{The {\it Herschel} Space Observatory view of dust in M81}

\author{G. J. Bendo\inst{1}, 
    C. D. Wilson\inst{2}, 
    M. Pohlen\inst{3},
    M. Sauvage\inst{4}, 
    R. Auld\inst{3},
    M. Baes\inst{5},
    M. J. Barlow\inst{6},
    J. J. Bock\inst{7},
    A. Boselli\inst{8},
    M. Bradford\inst{7},
    V. Buat\inst{8},
    N. Castro-Rodriguez\inst{9},
    P. Chanial\inst{4},
    S. Charlot\inst{10},
    L. Ciesla\inst{8},
    D. L. Clements\inst{1},
    A. Cooray\inst{11},
    D. Cormier\inst{4},
    L. Cortese\inst{3},
    J. I. Davies\inst{3},
    E. Dwek\inst{12},
    S. A. Eales\inst{3},
    D. Elbaz\inst{4},
    M. Galametz\inst{4},
    F. Galliano\inst{4},
    W. K. Gear\inst{3},
    J. Glenn\inst{13},
    H. L. Gomez\inst{3},
    M. Griffin\inst{3},
    S. Hony\inst{4},
    K. G. Isaak\inst{14},
    L. R. Levenson\inst{7},
    N. Lu\inst{7},
    S. Madden\inst{4},
    B. O'Halloran\inst{1},
    K. Okumura\inst{4},
    S. Oliver\inst{15},
    M. J. Page\inst{16},
    P. Panuzzo\inst{4},
    A. Papageorgiou\inst{3},
    T. J. Parkin\inst{2},
    I. Perez-Fournon\inst{9},
    N. Rangwala\inst{13},
    E. E. Rigby\inst{17},
    H. Roussel\inst{10},
    A. Rykala\inst{3},
    N. Sacchi\inst{18},
    B. Schulz\inst{19},
    M. R. P. Schirm\inst{2},
    M. W. L. Smith\inst{3},
    L. Spinoglio\inst{18},
    J. A. Stevens\inst{20},
    S. Sundar\inst{10},
    M. Symeonidis\inst{16},
    M. Trichas\inst{1},
    M. Vaccari\inst{21},
    L. Vigroux\inst{10},
    H. Wozniak\inst{22},
    G. S. Wright\inst{23},
    W. W. Zeilinger\inst{24}
}
\authorrunning{G. J. Bendo et al.}

\institute{Astrophysics Group, Imperial College, Blackett Laboratory, Prince 
        Consort Road, London SW7 2AZ, UK 
        \email{g.bendo@imperial.ac.uk}
    \and
        Dept. of Physics \& Astronomy, McMaster University, Hamilton,  
        Ontario, L8S 4M1, Canada
    \and
        School of Physics and Astronomy, Cardiff University, Queens  
        Buildings The Parade, Cardiff CF24 3AA, UK
    \and
        Laboratoire AIM, CEA/DSM - CNRS - Universit\'e Paris Diderot, 
        Irfu/Service d'Astrophysique, 91191 Gif sur Yvette, France
    \and	
        Sterrenkundig Observatorium, Universiteit Gent, Krijgslaan 281 S9,  
        B-9000 Gent, Belgium
    \and
        Department of Physics and Astronomy, University College London,  
        Gower Street, London WC1E 6BT, UK
    \and
        Jet Propulsion Laboratory, Pasadena, CA 91109, United States;  
        Department of Astronomy, California Institute of Technology, Pasadena,  
        CA 91125, USA
    \and
        Laboratoire d'Astrophysique de Marseille, UMR6110 CNRS, 38 rue F.  
        Joliot-Curie, F-13388 Marseille France
    \and
        Instituto de Astrofísica de Canarias, C/V\'ia L\'actea s/n, E-38200 La  
        Laguna, Spain
    \and
        Institut d'Astrophysique de Paris, UMR7095 CNRS, Universit\'e Pierre  
        \& Marie Curie, 98 bis Boulevard Arago, F-75014 Paris, 
        France
    \and
        Department of Physics \& Astronomy, University of California, Irvine,
        CA 92697, USA
    \and
        Observational  Cosmology Lab, Code 665, NASA Goddard Space Flight   
        Center Greenbelt, MD 20771, USA
    \and
        Department of Astrophysical \& Planetary Sciences, CASA CB-389,  
        University of Colorado, Boulder, CO 80309, USA
    \and
        ESA Astrophysics Missions Division, ESTEC, PO Box 299, 2200 AG
        Noordwijk, The Netherlands
    \and
        Astronomy Centre, Department of Physics and Astronomy, University of  
        Sussex, UK
    \and
        Mullard Space Science Laboratory, University College London,  
        Holmbury St Mary, Dorking, Surrey RH5 6NT, UK
    \and
        School of Physics \& Astronomy, University of Nottingham, University  
        Park, Nottingham NG7 2RD, UK
    \and
        Istituto di Fisica dello Spazio Interplanetario, INAF, Via del Fosso  
        del Cavaliere 100, I-00133 Roma, Italy
    \and
        Infrared Processing and Analysis Center, California Institute of  
        Technology, Mail Code 100-22, 770 South Wilson Av, Pasadena, CA 91125,  
        USA
    \and
        Centre for Astrophysics Research, Science and Technology Research  
        Centre, University of Hertfordshire, College Lane, Herts AL10 9AB, 
        UK
    \and
        University of Padova, Department of Astronomy, Vicolo Osservatorio  
        3, I-35122 Padova, Italy
    \and
        Observatoire Astronomique de Strasbourg, UMR 7550 Universit\'e de  
        Strasbourg - CNRS, 11, rue de l'Universit\'e, F-67000 
        Strasbourg
    \and
        UK Astronomy Technology Center, Royal Observatory Edinburgh, 
        Edinburgh, EH9 3HJ, UK
    \and
        Institut f\"ur Astronomie, Universit\"at Wien, T\"urkenschanzstr. 17,  
        A-1180 Wien, Austria
}

\abstract{We use {\it Herschel} Space Observatory data to place
  observational constraints on the peak and Rayleigh-Jeans slope of
  dust emission observed at 70-500~$\mu$m in the nearby spiral galaxy
  \object{M81}.  We find that the ratios of wave bands between 160 and
  500~$\mu$m are primarily dependent on radius but that the ratio of
  70 to 160~$\mu$m emission shows no clear dependence on surface
  brightness or radius.  These results along with analyses of the
  spectral energy distributions imply that the 160-500~$\mu$m emission
  traces 15-30~K dust heated by evolved stars in the bulge and disc
  whereas the 70~$\mu$m emission includes dust heated by the active
  galactic nucleus and young stars in star forming regions.}


\maketitle

\section{Introduction}

The {\it Herschel} Space Observatory\footnote{{\it Herschel} is an ESA space
  observatory with science instruments provided by Principal
  Investigator consortia. It is open for proposals for observing time
  from the worldwide astronomical community. } \citep{pilbrattetal10}
provides an unprecedented view of the far-infrared and submillimetre
emission from nearby galaxies.  At wavelengths of 70-160 $\mu$m, the
PACS instrument \citep{poglitschetal10} can produce images with
resolutions of $6^{\prime\prime}$-$12^{\prime\prime}$ that are
  superior to what can be achieved with the {\it Spitzer} Space
  Telescope.  At 250-500~$\mu$m, the SPIRE instrument 
  \citep{getal10} produces images with unprecedented sensitivities to
diffuse and point-like submillimetre emission.  We can use these data
to construct spectral energy distributions (SEDs) that sample
the peak and Rayleigh-Jeans side of thermal dust emission, thus
allowing us to probe the coldest dust components in nearby galaxies
and place superior constraints on dust temperatures and masses.  As
part of the Very Nearby Galaxies Survey (VNGS), we have imaged the
spiral galaxy M81 (NGC 3031) at 70, 160, 250, 350, and 500~$\mu$m with
PACS and SPIRE.  M81 is a nearby \citep[$3.63\pm0.13$~Mpc;][]{fetal01}
SA(s)ab \citep{ddcbpf91} galaxy at an inclination of $59.0^\circ$
\citep{detal08} with well defined spiral arms.  The {\it Herschel} data
allow us to extract SEDs for $\sim0.7$~kpc subregions that are small
enough that we can distinguish arm and interarm regions within M81.
We use these data to explore the dust temperatures and masses
and to understand the heating sources for the dust.

\section{Observations and data reduction}
\label{s_data}

The PACS observations were performed as four pairs of orthogonal scans
covering $40^\prime\times40^\prime$ using a
20$^{\prime\prime}$~s$^{-1}$ scan rate. PACS can perform simultaneous
observations in only two wave bands; we chose the 70 and 160 $\mu$m
bands since they were expected to bracket the peak of the SED better.
The data were reduced using a combination of an adapted {\it Herschel}
Interactive Processing Environment (HIPE) 3.0 pipeline and
Scanamorphos (Roussel et al. in prep.).  Starting from the raw detector
timelines, HIPE was used to mask dead and saturated pixels, convert
the signal to Jy pixel$^{-1}$, and remove cosmic rays.  Scanamorphos
was then used to map the data while simultaneously removing $1/f$
drifts in the signals.  Finally, we subtracted the median backgrounds
from the images.  The photometric calibration has an accuracy of 10\%
at 70 $\mu$m and 20\% at 160 $\mu$m, and the full-width half-maxima
(FWHM) of the 70 and 160~$\mu$m point spread functions (PSFs) are
$6^{\prime\prime}$ and $12^{\prime\prime}$, respectively
\citep{poglitschetal10}.  The RMS noise levels are 0.12 mJy
arcsec$^{-2}$ in both the 70 and 160~$\mu$m bands.

\begin{figure*}
    \centering
    \includegraphics[height=6cm]{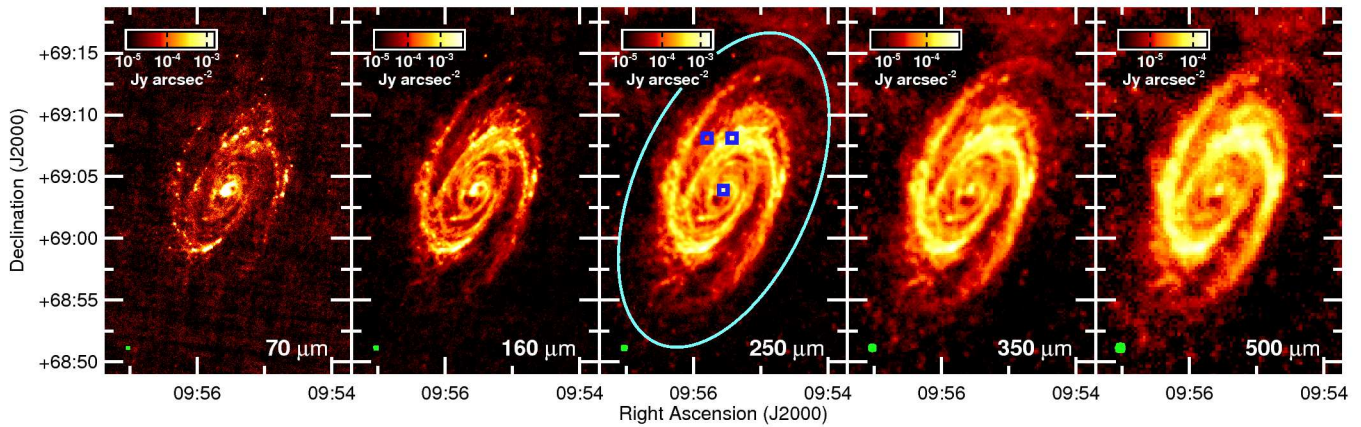}
    \caption[width=\textwidth]{70-500~$\mu$m images of M81 covering
      $20^\prime \times 30^\prime$ with north up and east to the left.
      The images are scaled logarithmically.  The green circles in the
      lower left corner of each image show the FWHM of the PSF.  The
      cyan ellipse in the 250~$\mu$m image shows the $D_{25}$ isophote
      \citep[$26^{\prime}.9\times14^{\prime}.1$;][]{ddcbpf91}; the
      radius is equivalent to 14 kpc.  The blue squares in the
      250~$\mu$m image show the $42^{\prime\prime}$ regions for which
      SEDs are plotted in Fig.~\ref{f_sed}.}
    \label{f_img}
\end{figure*}

\begin{figure*}
    \centering
    \includegraphics[height=6cm]{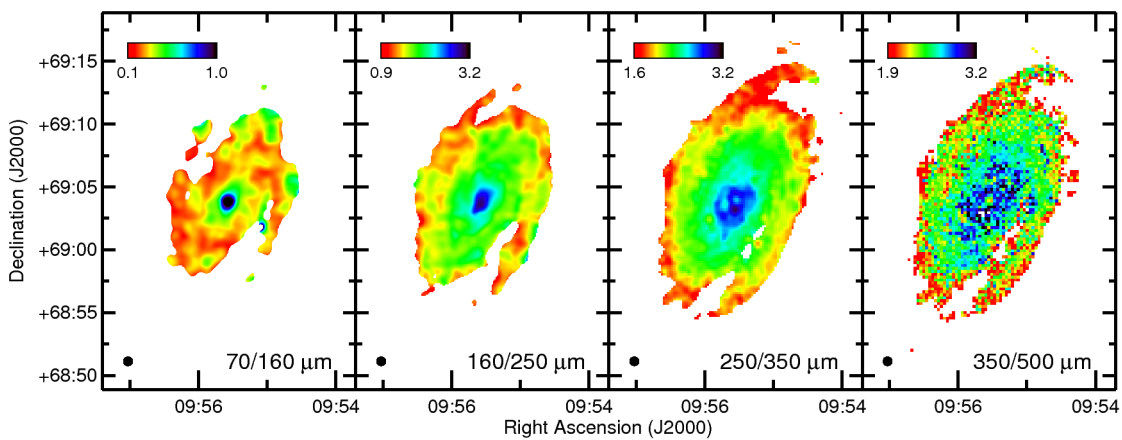}
    \caption[width=\textwidth]{Images of the 70/160, 160/250, 250/350,
      and 350/500~$\mu$m surface brightness ratios in the optical disc
      of M81.  Each image is created using data with PSFs that match
      the 500~$\mu$m PSF.  Regions not detected at the $3\sigma$ level
      in the two bands used for each ratio are left blank.  The image
      sizes and orientations are the same as for Fig.~\ref{f_img}.
      The circles in the lower left corner of each image show the FWHM
      of the 500~$\mu$m PSF.}
    \label{f_img_ratio}
\end{figure*}

The SPIRE observations were performed as two orthogonal scans covering
$40^\prime\times40^\prime$ using a 30$^{\prime\prime}$~s$^{-1}$ scan
rate.  A modified HIPE 3.0 detector timeline pipeline was used to
remove cosmic rays, flux calibrate the data, and apply temperature
drift and response corrections \citep[see][for details]{pohlenetal10}.
We then removed offsets between the detector timelines in two steps.
First, we subtracted the median signal from each bolometer observed
during the entire observation.  Then we applied an iterative process
to remove residual baseline signals that appear as stripes in the
maps.  In this process, we first created a map.  Then, for each
bolometer timeline in each scan leg, we measured the signal in the map
that corresponded to the bolometer's position, we calculated the
median difference between the bolometer signal and the corresponding
map signal, and we subtracted this function from the bolometer signal.
These steps were repeated 40 times to completely remove stripes from
the data.  Finally, we subtracted median background signals from the
images.  The resulting images have flux calibration uncertainties of
15\%, and the 250, 350, and 500~$\mu$m PSFs have FWHM of
$18^{\prime\prime}$, $25^{\prime\prime}$, and $37^{\prime\prime}$,
respectively \citep{swinyardetal10}.  The RMS noise levels are 0.040,
0.019, and 0.008 mJy arcsec$^{-2}$ in the 250, 350, and 500~$\mu$m
bands, respectively.

To create ratios of surface brightnesses measured in two wave bands,
we matched the PSFs, which we treated as Gaussian, to the PSF of the
500~$\mu$m data.  For statistical analyses on surface brightness
ratios and for creating SEDs of subregions within the galaxies, we
then rebinned the data in all images into $42^{\prime\prime}$
($\sim0.7$~kpc) square pixels (selected because it is an integer
multiple of the 500~$\mu$m pixel size that is larger than the PSF FWHM
for the 500~$\mu$m data).  For these analyses, we only used used
$42^{\prime\prime}$ pixels with $3\sigma$ detections in all
bands.

\section{Results}

Figure~\ref{f_img} shows the structures traced by the various {\it Herschel}
wave bands, which look very similar to each other and to the 5.7-24~$\mu$m
{\it Spitzer} images \citep{getal04, wetal04}.  All images trace the
same spiral structure and individual infrared sources within the disc
of the galaxy.  Diffuse, extended sources detected outside the optical
disc of M81 in the {\it Herschel} data seem most likely to be associated
with dust in the Milky Way (Davies et al. in prep.).

\begin{figure}
    \centering
    \includegraphics{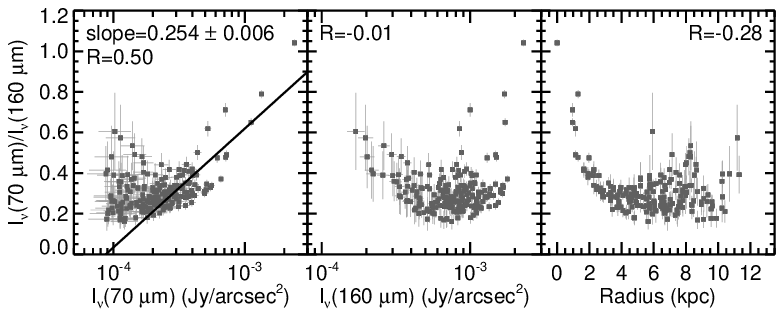}
    \includegraphics{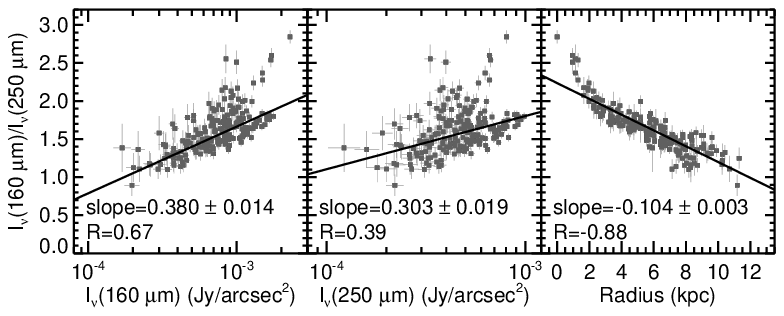}
    \includegraphics{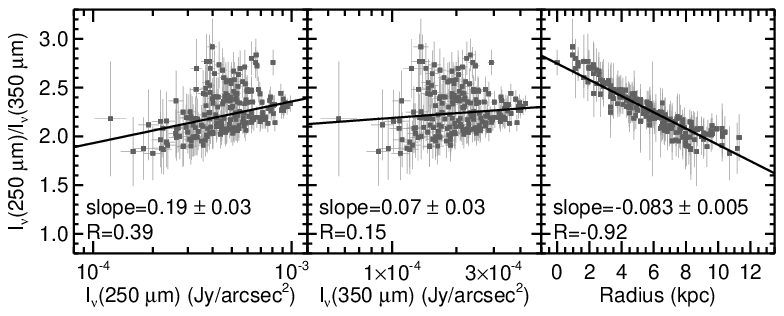}
    \includegraphics{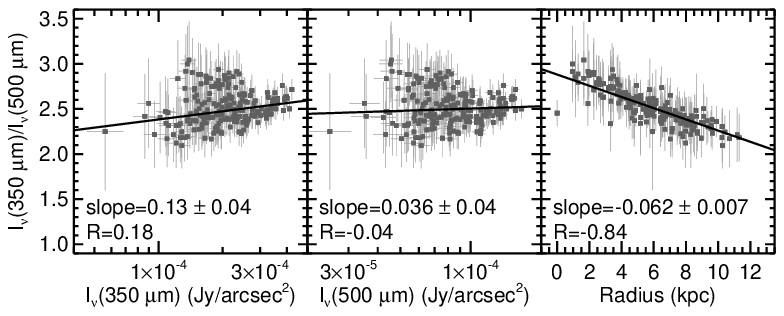}
    \caption[width=\textwidth]{The 70/160, 160/250, 250/350, and
      350/500~$\mu$m surface brightness ratios versus the surface
      brightness (left and center) and inclination-corrected
      galactocentric radius (right).  The data were measured in
      $42^{\prime\prime}$ subregions in images with PSFs that matched
      the PSF of the 500~$\mu$m data.  Best fit lines are shown for
      all plotted data except for two relations involving the
      70/160~$\mu$m ratio, where the fits were very poor;
      corresponding slopes are given in the panels.  The $R$ values
      are the Pearson correlation coefficients for the plotted data.
      Note that, in the left-side and center panels, the logarithm of
      the surface brightnesses are used for the best fit lines and
      correlation coefficients.}
    \label{f_ratiovar}
\end{figure}

To understand the heating mechanism for the dust, we examined how
surface brightness ratios varied with surface brightness and with
radius.  Variations with surface brightness would suggest that the
dust is heated locally and that the emission is linked to star
formation, whereas radial variations in the ratios would indicate that
the dust emission is more strongly affected by the evolved stellar
populations in the bulge and disc.  Figure~\ref{f_img_ratio} shows
images of the 70/160, 160/250, 250/350, and 350/500~$\mu$m surface
brightness ratios.  Additionally, Fig.~\ref{f_ratiovar} shows how
the ratios measured in $42^{\prime\prime}$ ($\sim0.7$~kpc) subregions
vary with surface brightness and galactocentric radius.
 
The absolute value of the correlation coefficient $R$ for the
relations between radius and either the 160/250, 250/350, or
350/500~$\mu$m ratios is generally higher than that for the relations
between surface brightness and the ratios, which shows that these
ratios are more strongly dependent on radius (although the
160/250~$\mu$m ratio may also be partly dependent on 160~$\mu$m
surface brightness based on the high value of $R$).  This is
consistent with the weak or absent infrared-bright regions or spiral
structure in the images of the 160/250, 250/350, and 350/500~$\mu$m
ratios.  Moreover, the $R^2$ values, which equal the fraction of the
variance in the data that can be accounted for by the best fit line,
indicate that $>$70\% of the variance in the 160/250, 250/350, and
350/500~$\mu$m ratios can be accounted for by the relation with
radius.

In contrast, the 70/160~$\mu$m ratio does not vary monotonically with
radius except within 2~kpc, a region in which the gradient in
the 160/250~$\mu$m ratio also increases.  This could represent
enhanced dust heating within this radius that is powered by the active
galactic nucleus (AGN), by strong central star formation activity, or
by the bulge stars, which have a high central density.  The
70/160~$\mu$m ratio versus 160~$\mu$m surface brightness exhibits no
obvious trend and only a statistically weak trend (with $R^2<0.3$) is
visible in the plot of the 70/160~$\mu$m ratio versus 70~$\mu$m
surface brightness, although the best fit line poorly describes the
data.  None the less, Fig.~\ref{f_img_ratio} shows that the ratio
increases to $\gtrsim0.3$ in infrared-bright regions in the spiral
arms.

\begin{figure}
    \centering
    \includegraphics{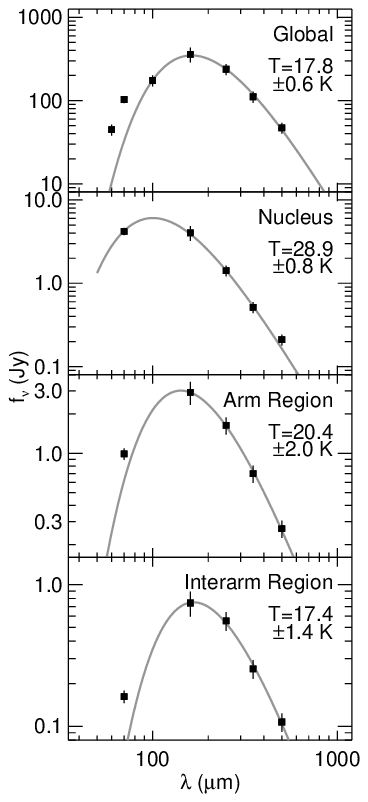}
    \includegraphics{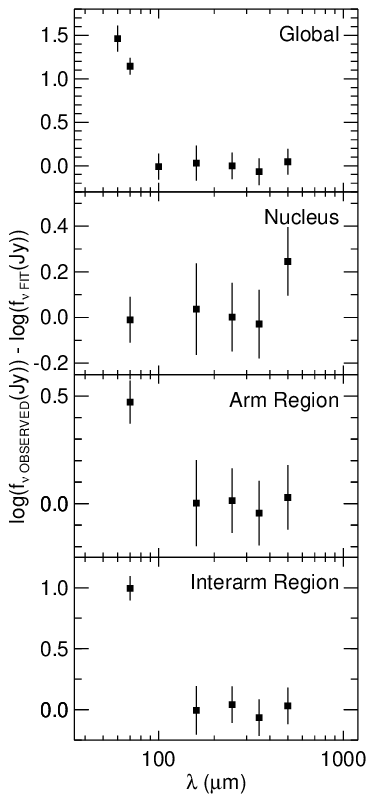}
    \caption[width=\textwidth]{On the left, the global SED as well as
      the SEDs for the three $42^{\prime\prime}$ ($\sim 0.7$~kpc)
      regions shown in Fig.~\ref{f_img}.  The SEDs for the subregions
      were measured in data with PSFs that matched to the PSF of the
      500~$\mu$m data.  The grey line is the blackbody modified with a
      $\lambda^{-2}$ emissivity function fit to the data.  On the
      right are the residuals from the fit in logarithm space.}
    \label{f_sed}
\end{figure}

Figure~\ref{f_sed} shows the SED integrated across the optical disc
(with supplemental 60 and 100~$\mu$m IRAS data added from
\citet{retal88}) as well as the SEDs for the $42^{\prime\prime}$
regions centered on the nucleus, and examples of an infrared-bright
source and an interarm region.  Based on visually inspecting and
fitting functions to the SEDs, these example regions were typical to
similar regions at similar radii.  In the nucleus, we were able to fit
a single blackbody modified with a $\lambda^{-2}$ emissivity function
(based on the \citet{ld01} emissivity function) to the 70-350~$\mu$m
data, but the 500~$\mu$m data point could not be fit with the same
thermal component, although the mismatch between the fit and model is
only $2\sigma$.  This result and the low 350/500~$\mu$m ratio for the
nucleus seen in Fig.~\ref{f_ratiovar} (which is $3\sigma$ below the
best fit line) suggest that the 500~$\mu$m nuclear emission likely
includes a non-thermal component associated with the AGN.  Based on
the SED fit, we estimate that the non-thermal 500~$\mu$m emission is
$0.05 \pm 0.03$~Jy, which is $\sim2.5\times$ below a power law
extrapolated from the mm and cm data presented by \citet{metal08}.
The discrepancy could be explained by the low signal-to-noise in the
estimate from the SED fit or by variability in the AGN emission; the
870~$\mu$m flux density has been observed to vary by $3\times$
\citep{metal08}.

In the other SEDs, we found that single blackbodies modified with
$\lambda^{-2}$ emissivity functions could be fit accurately to the
$>$100~$\mu$m data without the fit overpredicting the observed
70~$\mu$m measurement, but fits that included the 70~$\mu$m data point
did not accurately replicate the peak of the SED.  No evidence is
found for the excess emission at submillimetre wavelengths sometimes
attributed to dust with $<10$~K temperatures or shallow emissivities,
although prior results had indicated that this emission would be more
prominent at $>$500~$\mu$m \citep[e.g.][]{gmjwb05, betal06, zpxkl09,
  oetal10}.  By applying to the data for the global SED the equation
$M_{dust}=[f_\nu D^2]/[\kappa B(\nu,T)]$ (where $D$ is distance,
$\kappa_\nu$ is the dust opacity from \citet{ld01}, and $B(\nu,T)$ is
the best fitting modified blackbody), we estimated the global dust
mass to be $3.4 \pm 0.5 \times10^7$~M$_\odot$.  Given that the atomic
gas mass is $3.64 \pm 0.18\times10^9$~M$_\odot$ \citep{wetal08} and
the molecular gas mass is negligible in comparison
\citep[][S\'anchez-Gallego et al. in prep.]{s93}, we estimate that the
gas-to-dust ratio is $107 \pm 17$, which is within the range of
$\sim100$-200 expected for solar metallicity objects based on the
depletion of metals from the gaseous phase of the interstellar medium
or comparisons of gas column density to dust extinction
\citep[e.g][]{w03}.  Hence, this simplistic modified blackbody fit may
be a fair representation of the emission from the bulk of the dust
mass in M81, although more sophisticated modeling should not only
yield more accurate masses but also describe the emission from warmer
dust components.

The SED fits along with the results from Figs.~\ref{f_img_ratio} and
\ref{f_ratiovar} imply that the 70~$\mu$m band traces dust heated by a
different source than the dust that primarily emits in the
160-500~$\mu$m bands.  Although the 70/160~$\mu$m ratio exhibits a lot
of scatter, the enhancements in the 70/160~$\mu$m ratio in the spiral
arms implies that the 70~$\mu$m band may be affected by star formation
on local scales.  Meanwhile, the radial variations in the
160-500~$\mu$m ratios and the SED fits suggest that $\sim20$\% of the
60~$\mu$m emission, $\sim30$\% of the 70~$\mu$m emission, and
$\sim100$\% of the $>100~\mu$m emission originates from dust heated by
evolved disc and bulge stars.  This is consistent with prior results
suggesting that $\sim 5$-100\% of the 60 and 100~$\mu$m emission from
nearby galaxies originates from dust heated by evolved stars
\citep[e.g.][]{st92,wg96}.  If this interpretation is correct, we
anticipate that dust emitting at 160-500~$\mu$m in other galaxies with
relatively large fractions of old stars (E-Sab galaxies) will also
have 160-500~$\mu$m colours that depend upon radius, but galaxies with
relatively large fractions of young stars (Sc-Im galaxies) will have
160-500~$\mu$m colours that may depend more on infrared surface
brightness, as heating by the evolved stellar population becomes
insignificant.  The results also imply that the conversion of infrared
fluxes integrated over very broad ranges (e.g. 8-1000~$\mu$m) to star
formation rates, as done by \citet{zwcl08}, \citet{retal09}, and
\citet{ketal09}, will be accurate as long as the integrals contain a
significant amount of emission shortward of 160~$\mu$m that traces
dust heated by star formation.  However, it may not be possible to
derive accurate star formation rates from dust emission measured
solely at $>$160~$\mu$m.

In conclusion, these results for M81 demonstrate how {\it Herschel}
70-500~$\mu$m data can be used to not only measure more accurate dust
temperatures and masses but also determine the dust heating sources in
nearby galaxies.  Further work with data from the VNGS and other
surveys will allow us to determine whether dust traced by the
160-500~$\mu$m bands in other spiral galaxies is also heated by
evolved stellar populations and whether variations in the relative
strength of dust heating by evolved stars varies across the Hubble
sequence.

\begin{acknowledgements}
We thank A. Fraceschini and E. Murphy for comments on this paper.
SPIRE has been developed by a consortium of institutes led by Cardiff
Univ. (UK) and including Univ. Lethbridge (Canada); NAOC (China); CEA,
LAM (France); IFSI, Univ. Padua (Italy); IAC (Spain); Stockholm
Observatory (Sweden); Imperial College London, RAL, UCL-MSSL, UKATC,
Univ. Sussex (UK); Caltech, JPL, NHSC, Univ. Colorado (USA). This
development has been supported by national funding agencies: CSA
(Canada); NAOC (China); CEA, CNES, CNRS (France); ASI (Italy); MCINN
(Spain); SNSB (Sweden); STFC (UK); and NASA (USA).  PACS has been
developed by a consortium of institutes led by MPE (Germany) and
including UVIE (Austria); KUL, CSL, IMEC (Belgium); CEA, OAMP
(France); MPIA (Germany); IFSI, OAP/AOT, OAA/CAISMI, LENS, SISSA
(Italy); IAC (Spain). This development has been supported by the
funding agencies BMVIT (Austria), ESA-PRODEX (Belgium), CEA/CNES
(France), DLR (Germany), ASI (Italy), and CICT/MCT (Spain).

\end{acknowledgements}

\bibliographystyle{aa}

\end{document}